\documentstyle[preprint,prl,aps]{revtex}

\begin{document}

\title{Dynamical Jahn-Teller Effect and Berry Phase
in Positively Charged Fullerene I. Basic Considerations}
\author{Paolo De Los Rios,$^{1,2}$\thanks{Email: delos@hp720.cm.sissa.it},
Nicola Manini,$^{1,2}$\thanks{Email: manini@sissa.it} and
Erio Tosatti$^{1,2,3}$\thanks{Email: tosatti@sissa.it}}
\address{$^1$ Istituto Nazionale di Fisica della Materia (INFM)}
\address{$^2$ International School for Advanced Studies (SISSA), Via Beirut 4,
I-34013 Trieste, Italy}
\address{$^3$ International Centre for Theoretical Physics (ICTP), P.O. Box
586, I-34014 Trieste, Italy}
\date{\today}
\maketitle

\begin{abstract}
We study the Jahn-Teller effect of positive fullerene ions $^2$C$_{60}^{+}$
and $^1$C$_{60}^{2+}$. The aim is to discover if this case, in analogy with
the negative ion, possesses a Berry phase or not, and what are the
consequences on dynamical Jahn-Teller quantization. Working in the linear
and spherical approximation, we find no Berry phase in $^1$C$_{60}^{2+}$,
and presence/absence of Berry phase for coupling of one $L=2$ hole to an
$L=4$/$L=2$ vibration. We study in particular the special equal-coupling
case ($g_2=g_4$), which is reduced to the motion of a particle on a
5-dimensional sphere. In the icosahedral molecule, the final outcome
assesses the presence/absence of a Berry phase of $\pi$ for the $h_u$ hole
coupled to $G_g$/$H_h$ vibrations. Some qualitative consequences on
ground-state symmetry, low-lying excitations, and electron emission from
C$_{60}$ are spelled out.
\end{abstract}

\pacs{PACS numbers: 33.10.Lb,71.38.+i,74.70.Wz}

\section{Introduction}

Non-adiabatic systems, where the quantum mechanics of electrons and that of
ions are deeply entangled, are among the most fascinating in physics. In
condensed matter, entire classes of phenomena such as superconductivity or
charge-density waves are of this nature. Dynamical Jahn-Teller (DJT)
centers (molecules and ions, impurities, nuclei) constitute perhaps the
simplest ``zero-dimensional'' example of nonadiabatic behavior, and have
attracted attention for over half a century\cite{Englman}.

An amusing anomaly has long been pointed out\cite{lhOP,lh,OP} in certain
simple DJT centers, where an $e$ electronic doublet is Jahn-Teller (JT)
coupled to a vibrational (``pseudo-rotational'') $E$ doublet. In modern
language, this anomaly can be seen as a Berry phase
\cite{Berry,Delacretaz}. In simple words, what happens in these systems is
that there exists a class of closed paths (loops) in the $E$ coordinate
space, regarded as classical, adiabatically transporting the electronic
ground state from an initial value $\psi_0$ to a final value
$e^{i\gamma}\psi_0=-\psi_0$, instead of into $\psi_0$ itself. There is a
Berry phase $\gamma=\pi$, whose origin lies in the fact that the point of
electronic degeneracy (the ``conical point''\cite{Mead,lh}) is irreducibly
surrounded by the trajectory in vibron space.

This anomaly which is present at the adiabatic level, where the ionic
coordinated are classical, has very interesting consequences when the full
joint ion-electron problem is finally quantized. In the $e\otimes E$ case
cited above, there appears a new conserved ``pseudo-angular momentum''
quantum number, whose values are
half-integer\cite{Englman,Delacretaz}. Consequently, all vibronic levels
become twofold degenerate, as appropriate to a rotor coupled to a
pseudospin $\frac12$\cite{Englman,Delacretaz}.

It is an interesting question therefore, to inquire whether other nontrivial
cases of DJT are realized in nature, which also possess a Berry
phase. High-symmetry polyatomic molecules are an obvious place to look at.

The fullerene molecule, C$_{60}$, has a very high ${\cal I}_h$ symmetry,
and a nondegenerate ground state. Because of high symmetry, however,
fullerene ions, both negative and positive have a high orbital degeneracy,
and are therefore good candidates.

In our previous work, we have studied\cite{AMT,MTA} the DJT problem of
negatively-charged fullerene ions, where $1\leq n\leq 5$ electrons partly
fill the $t_{1u}$ Lowest Unoccupied Molecular Orbital (LUMO) of the neutral
molecule. This electronic state couples linearly to the fivefold degenerate
$H_g$ vibrations. For odd electron number $n$, the spectrum of this coupled
dynamical $t_{1u}\otimes H_g$ system is affected by anomalies which can
again be usefully described in terms of a Berry phase\cite{Berry}.

Among notable consequences, we have predicted the dominance
of $p$-wave attachment for low-energy electrons\cite{attachment}, a
negative pairing energy for odd $n$\cite{AMT,MTA}, characteristic
spectroscopic anomalies\cite{Reno} and a mechanism for kinematic electron
pairing\cite{MTD,SAMTP,SMPT}

 From the mathematical point of view, nevertheless, the DJT of C$_{60}^-$,
namely $t\otimes H$, was not particularly new: the analytical machinery was
basically the same as that previously studied by O'Brien\cite{ob69,ob71}
for the $t\otimes (E\oplus T)$ equal coupling case.

In the present paper, we move on to analyze the problem of the JT effect in
the {\em positively} charged fullerene ions C$_{60}^{n+}$, inquiring in
particular about the possible relevance of a Berry phase in these systems
too. As it will turn out, the answer is again affirmative (though with some
qualifications). Moreover, we shall deal in this case with the less trivial
JT problems $h\otimes H$ and $h\otimes (G\oplus H)$, where the search for
an adiabatic anomaly will now be done afresh.

Fullerene cations C$_{60}^{n+}$ are obtained by removing $n\leq 10$
electrons from (i.e. adding $n\leq 10$ holes to) the fivefold degenerate
$h_u$ Highest Occupied Molecular Orbital
(HOMO)\cite{Satpathy,Ozaki,Savina}.  Our approach allows us to treat the
one-hole case ($n=1$) as well as the two-holes ($n=2$) singlet case, in a
way which is essentially analytical, albeit for a very idealized set of
parameters.

A hole in the HOMO interacts with the C$_{60}$ vibrational modes. We focus
on modes which couple linearly with the hole density fluctuations (which in
turn are of course quadratic in the hole wave function). Hence, the
icosahedral symmetry of the molecule restricts the linearly coupled modes
to those of the same symmetry as the irreducible representation of the
group (${\cal I}_h$) contained in the symmetric product of a hole
representation with itself\cite{Cesare}:
\begin{equation}
\{ h_{u} \otimes h_{u} {\}}_s \approx a_g \oplus h_g \oplus (g_g \oplus h_g)
\label{couplingstructure}
\end{equation}

The ${\cal I}_h$ group is a subgroup of the full rotation group
O(3). Therefore, a general representation of O(3) decomposes into a finite
number of representations of ${\cal I}_h$\cite{Altmann}. In particular,
the $L=2$ ${\cal D}^{(2\pm)}$ representations correspond to $h_{g/u}$ in
icosahedral symmetry, and $L=4$ ${\cal D}^{(4\pm)}$ decomposes into $g_{g/u}
\oplus h_{g/u}$.  Equation (\ref{couplingstructure}) may be rewritten, in
O(3) notation,
\begin{equation}
\{ {\cal D}^{(2-)} \otimes {\cal D}^{(2-)}{\}}_s \approx
		{\cal D}^{(0+)} \oplus {\cal D}^{(2+)} \oplus {\cal D}^{(4+)}
\label{so3}
\end{equation}
In order to take advantage of spherical symmetry, and as a simplifying
approximation at this initial stage, we neglect the icosahedral details of
the molecular shape, and we study the hole-vibron coupling within the O(3)
scheme, as though C$_{60}$ was a perfect sphere\cite{CeulemansIII}.  We
study here therefore the problem of ${\cal D}^{(2-)}$ holes coupled to
${\cal D}^{(0+)}$, ${\cal D}^{(2+)}$ and ${\cal D}^{(4+)}$ vibrations.

The coupling with a ${\cal D}^{(0+)}$ vibration is a trivial polaronic
problem, solved exactly as a displaced oscillator, and is irrelevant here,
since it shifts energies, but does not cause splittings.  We concentrate on
quadrupolar (${\cal D}^{(2+)}$) and hexadecapolar (${\cal D}^{(4+)}$)
modes, which lead instead to a non-trivial JT effect. In order to simplify
the notation, we shall switch from the O(3) to the SO(3) notation, omitting
therefore the $\pm$ signs for the inversion.

The Hamiltonian we consider has the following structure:
\begin{equation}
H = H_0 + H_{h-v} \ , \ H_{h-v} = H_{h-v}^{(2)} + H_{h-v}^{(4)} ~ ,
\label{Hschematica}
\end{equation}
where $H_0$ describes a free (uncoupled) ${\cal D}^{(2)}$ hole and ${\cal
D}^{(2)}$, ${\cal D}^{(4)}$ vibrations, and $H_{h-v}$ the hole-vibron
interaction, namely
\begin{eqnarray}
	H_0 & = & \sum_{L = 2,4} \hbar {\omega}_L \sum_{m=-L}^{L}
		\left(b_{L,m}^{\dagger} b_{L,m} + \frac{1}{2}\right) +
		(\epsilon - \mu) \sum_{\sigma= \uparrow , \downarrow}
		\sum_{m=-2}^{2} c_{m,\sigma}^{\dagger} c_{m,\sigma} \ ,
		\nonumber\\
	H_{h-v}^{(L)} & = & g_L \frac{\hbar {\omega}_L}{2}
		\sum_{\sigma,m_1,m_2} (-1)^{m_2}
		\left<L,m_1 \mid 2,-m_2;2,m_1+m_2\right> \times
 \label{Hesplicita} \\
	 	& & \hspace{10 mm}
		\times \left[b_{L,m_1}^{\dagger} +(-1)^{m_1} b_{L,-m_1}\right]
		c_{m_2,\sigma}^{\dagger} c_{m_1+m_2,\sigma} ~ . \nonumber
\end{eqnarray}
Here, $\hbar \omega_L$ are the energies of the (harmonic) vibrations
$b_{L,m}^{\dagger}$, $g_L$ are dimensionless coupling parameters, $<L,m
\mid L_1,m_1;L_2,m_2>$ are Clebsch-Gordan coefficients, and there is no
loss of generality in choosing the energy zero so that $\epsilon =
\mu$. For simplicity, we shall assume here a single $L=2$ and a single
$L=4$ mode corresponding to ${\cal D}^{(2)}(=H_g)$ and ${\cal
D}^{(4)}(=G_g+H_g)$ vibrons respectively. In real C$_{60}$, the spectrum is
more complex (see Tab.\ \ref{ExperEner:table}), involving 8 $H_g$ and 6
$G_g$ modes.  Of these, only two $H_g$ modes, namely $H_g(1)$ and $H_g(4)$
are actually derived from a ${\cal D}^{(2)}$ spherical vibration, while
only two doublets, namely $G_g(1)\oplus H_g(2)$ and $G_g(4)\oplus H_g(6)$
are of $L=4$ origin\cite{CeulemansIII}. The other modes derived from
different angular momenta ($L=3,5,6,7$) do not belong to a complete
$L$-multiplet, and cannot be simply treated in spherical symmetry. Our work
is therefore incomplete and of purely qualitative value.  Nevertheless, we
expect some of the features of the present simplified analysis to be of
more general relevance.  For example, close pairs of modes, such as
$G_g(3)\oplus H_g(3)$, or $G_g(6)\oplus H_g(8)$, have no reason to behave
qualitatively differently from our ${\cal D}^{(4)}$ mode, in spite of their
different $L=6$ and $L=7$ origins. We will also find a ``conservation of
Berry phases'': if a JT-coupled problem has a Berry phase, the extension of
the vibron space with the addition of a mode that does not carry a Berry
phase, leaves intact the original Berry phase. In our case, we will show
that the Berry phase is coming exclusively from the $h_u\otimes G_g$
coupling, the addition of the $H_g$ modes changing the dynamical
properties, but not the topological ones.

According to the different relative values of the four parameters
($\omega_2,g_2,\omega_4,g_4$) in $H$, different regimes can be attained. In
particular, when the coupling is extremely strong ($g_L\to \infty$), the JT
distortion is large, higher order terms (neglected in (\ref{Hesplicita}))
will in principle become relevant. In such a case, the final state tends to
approach closely a static JT symmetry-broken distortion. When the coupling
is instead large but finite, there is a semiclassical region, where quantum
fluctuations are able to connect classical equivalent valleys {\em via}
tunneling, and so restore the full symmetry while also generating
characteristic low-energy tunneling excitations in the spectrum. The
intermediate coupling region $g_L\approx 1$ is, as usual, the less
predictable. However, the weak coupling limit ($g_L\to 0$) finally
simplifies again into a perturbative problem.  If either $g_4$ or $g_2$
alternatively vanish, we will have the special cases of ${\cal
D}^{(2)}\otimes {\cal D}^{(2)}$ or ${\cal D}^{(2)}\otimes {\cal D}^{(4)}$
JT respectively.  A special case we shall consider in greatest detail below
is that of equal coupling of degenerate oscillators $\omega_2=\omega_4$,
$g_2=g_4$.

This paper is organized as follows.  The weak-coupling region $g_2,g_4<<1$
is discussed in Sect.\ \ref{PerTh:sect}, obtaining approximate analytic
expressions for the splittings and shifts in the spectrum. Next, we shall
move to the opposite, strong-coupling case. Here, the classical JT minima
of the Born-Oppenheimer potential energy surfaces are studied first (Sect.\
\ref{topology:sect}); then, in Sect.\ \ref{semiclass:sect}, the ionic
kinetic energy is introduced semiclassically to obtain a description of the
low-energy quantum excitations in the strong-coupling limit. Exact
numerical diagonalizations will be used to shed light on a few particular
features of the intermediate-coupling regime. Finally, the relevance of
these preliminary theoretical results to the physics of fullerene cations
is discussed in Sect.\ \ref{discussion:sect}.

\section{The weak-coupling limit: perturbative approach}
\label{PerTh:sect}

The weak-coupling limit in a JT problem is a coupled situation in which
both the electronic and the vibronic degrees of freedom need to be treated
on equal footing, as equally ``fast''.  Their cooperative dynamics is
strictly regulated by quantum mechanics, and no ``classical degree of
freedom'' can be singled out in this limit, as opposed to the Berry-phase
semiclassical approach.  Nonetheless, the limit of small $g_2,g_4$ is
easily treated by perturbation theory, and the results are useful in
providing an exact match to the semiclassical spectrum on the opposite side
of the range of coupling.

The JT coupling (\ref{Hesplicita}) splits and shifts the harmonic levels to
second-order in $g_L$. We compute here these shifts for both the
fivefold-degenerate ground state and the 70-fold degenerate one-vibron
excitation in the 1-hole case and for equal frequencies
$\hbar\omega_2=\hbar\omega_4=1$.  As we have two small parameters, $g_2$
and $g_4$, and two different perturbing Hamiltonians, $H_{h-v}^{(2)}$ and
$H_{h-v}^{(4)}$, there is an infinite number of choices of the small
perturbing parameter, depending for example on the ratio $g_2/g_4$. We
compute the energy shifts for three special cases: $g_4=0$, i.e. perturbing
with respect to $H_{h-v}^{(2)}$, $g_2=0$, using $H_{h-v}^{(4)}$ alone, and
finally $g_2=g_4=g$, the equal coupling case.  The coefficients for the
three cases are collected in Tab.\ \ref{PerturbCoef:table}, along with the
appropriate SO(3) symmetry labels. The energy of a state (in units of
$\hbar\omega$) is thus given by
\begin{equation}
({\text{number\ of\ vibrons}}) + ({\text{coefficient\ in\ table}}) \times g^2 +
				{\cal O}(g^4) \ ,
\end{equation}
where the number of vibrons is one for all the states except for the ground
state at the first line.

We note on the second column five states, in the 1-vibron multiplet, whose
shifts are $-\frac14 g^2_2$, i.e. the same as for the ground state. These
states are readily identified as the 45-fold degenerate state corresponding
to the ${\cal D}^{(4)}$ vibron fundamental state (here uncoupled, since
$g_4=0$). Parallel considerations apply to the uncoupled ${\cal D}^{(2)}$
vibron in the third column, recognizable by the shifts of
$-\frac{9}{20}g^2_4$.

The equal-coupling case is interesting from the point of view of
symmetry. Pooler studied the Hamiltonian (\ref{Hesplicita}) from an
algebraic point of view\cite{Pooler}, discovering that the equal
frequencies -- equal coupling case $\omega_2=\omega_4$, $g_2=g_4$, is
actually characterized not only by the obvious SO(3) symmetry, but also by
rotation symmetry in a 5-dimensional space (SO(5) group). In the language
of the SO(5) irreducible representations, characterized by two quantum
numbers [$l$,$m$]\cite{Hamermesh}, the ${\cal D}^{(2)}$ hole is classified
as a [1,0] irreducible representation of SO(5), while the 14-fold
degenerate vibration (${\cal D}^{(2)}\oplus {\cal D}^{(4)}$) is a
realization of the [2,0] representation of SO(5)\cite{SO5:nota}.  In the
fourth column of Table\ \ref{PerturbCoef:table}, we note that the
one-vibron level in the equal-coupling case splits into three levels with
degeneracies 30, 35 and 5 respectively. The levels of degeneracy 5 and 30
are readily seen as representations [1,0] and [3,0] of
SO(5)\cite{Cornwell}. The 35-fold degenerate level, by table 10-3 of
reference\cite{Hamermesh} is identified with a [1,2] representation, as
indicated in the last column of Table\ \ref{PerturbCoef:table}. As this
large symmetry is an exact feature of the Hamiltonian, it is not limited to
the perturbative regime, but will be carried along to all values of
(equal-) coupling, and to the semiclassical limit (Sect.\
\ref{semiclass:sect}) in particular.

To conclude, we note that, while the $L=0,1,5,6$ levels are exactly the same
in the three perturbative regimes summarized in Tab.\
\ref{PerturbCoef:table}, the $L=2,3,4$ levels (marked in Tab.\
\ref{PerturbCoef:table} with symbols $^\dagger,^\ddagger,^*$), in the three
cases, are instead different linear combinations of the same states: this
fact has a very simple explanation.  The $L=0,1$ levels uniquely come from
the coupling of the ${\cal D}^{(2)}$ hole to the ${\cal D}^{(2)}$ vibron,
and the $L=5,6$ levels derive from the coupling of the ${\cal D}^{(2)}$
hole to the ${\cal D}^{(4)}$ vibron. The $L=2,3,4$ levels can come from
either of the previous couplings, and there is no {\em a priori} reason why
they should not mix among themselves.

\section{Topology of the Jahn-Teller manifold and the Berry phases}
\label{topology:sect}

In this section we want to understand if the JT hole-vibron coupling in
C$_{60}$ positive ions possesses a Berry phase or not. In order to do that,
the canonical steps are the following: (i) find the manifold, in the space
of classical ionic coordinates, formed by the set of equivalent minima of
Hamiltonian (\ref{Hesplicita}) (the minima of the Born-Oppenheimer (BO)
energy surface); (ii) study the adiabatic transport of the electronic
ground state wave-function -- in this case a unit vector spanning a
five-dimensional sphere -- while the ionic coordinates are taken along a
closed path on their manifold.  Representation (\ref{Hesplicita}) of
Hamiltonian (\ref{Hschematica}) is inconvenient for this purpose, and we
first perform a transformation of the electronic and vibronic operators to
a suitable representation.

\subsection{The real representation}

We start defining a vector of hole-destruction operators,
\begin{equation}
C_\sigma =
   (c_{2\sigma}, c_{1\sigma}, c_{0\sigma}, c_{-1\sigma}, c_{-2\sigma})^T ~ .
\label{vectc}
\end{equation}
In terms of this vector, the interaction Hamiltonian can be written as
\begin{equation}
H_{h-v}^{(L)} = \sum_\sigma C_\sigma^{\dag} \cdot B_L \cdot C_\sigma
\end{equation}
where $C_\sigma^{\dag}$ indicates the adjoint transpose of (\ref{vectc}),
and $B_2$ and $B_4$ are $5\times 5$ matrices whose elements are
combinations of vibrational coordinates.  Next, we apply the following
transformation to the electronic operators
\begin{equation}
\tilde{C_\sigma} = \left( \begin{array}{c} \tilde c_{1\sigma} \\
		\tilde c_{2\sigma}\\
		\tilde c_{3\sigma} \\
		\tilde c_{4\sigma} \\
		\tilde c_{5\sigma} \end{array}  \right) =
\frac{1}{\sqrt{2}} \left( \begin{array}{ccccc}
       i	& 0	& 0	& 0	& i   \\
       1	& 0	& 0	& 0	& -1   \\
       0	& 0	&\sqrt{2}i& 0	& 0   \\
       0	& -i	& 0	& i	& 0  \\
       0	& 1	& 0	& -1	& 0
			\end{array}  \right)  C_\sigma .
\label{c_transf}
\end{equation}
The transformed vector $\tilde{C_\sigma}$ contains the phase factors
necessary to give a suitable symmetry to the transformed $B$-matrices.

The final step converts the vibron operators into their real coordinate
representation:
\begin{equation}
q_{L,m} =
	\sum_{\lambda = -L}^{L} M_{m,\lambda} \left( b_{L,\lambda}^{\dag} +
         (-1)^{\lambda} b_{L,-\lambda} \right),
\label{b_transf}
\end{equation}
where
\begin{eqnarray} M_{m,\lambda \neq 0} & = & (2 \text{sign}(m))^{-\frac{1}{2}}
(\delta_{m,\lambda} + (-1)^m \text{sign}(m) \delta_{m,-\lambda}) \ ,\nonumber\\
M_{m,0} & = & {\delta}_{m,0} ~ ,
\label{def_M}
\end{eqnarray}
and the masses of both oscillators are taken as unity.

We rewrite $H_0$ and $H_{h-v}$ after the final transformation as
\begin{eqnarray}
H_0 &=& \sum_{L=2,4} \frac{\hbar \omega_L}{2} \sum_{l=-L}^{L}
		(-\partial_{L,l}^2 + q_{L,l}^2) \label{H0}\\
H_{h-v}^{(L)} &=& g_L \frac{\hbar {\omega}_L}{2}
	\sum_\sigma \tilde{C_\sigma}^\dagger \tilde B_L
	\tilde{C_\sigma} ~ ,	\label{HL}
\end{eqnarray}
where the matrices $\tilde B_2$ and $\tilde B_4$, the transformed of
$B_2$ and $B_4$ under (\ref{c_transf}), expressed in terms of $q_{L,m}$
(\ref{b_transf}) read
\begin{equation}
\tilde B_2 = \frac{1}{\sqrt{7}} \left( \begin{array}{ccccc}
	2 q_{2,0} & 0 & 2 q_{2,2} & \sqrt{3} q_{2,-1} & -\sqrt{3} q_{2,1} \\
	0 & 2 q_{2,0} & 2 q_{2,-2} & -\sqrt{3} q_{2,1} & -\sqrt{3} q_{2,-1} \\
	2 q_{2,2} & 2 q_{2,-2} & -2 q_{2,0} & -q_{2,-1} & -q_{2,1} \\
	\sqrt{3} q_{2,-1} & -\sqrt{3} q_{2,1} & -q_{2,-1} &  -q_{2,0} +
				\sqrt{3} q_{2,2} & -\sqrt{3} q_{2,-2} \\
	-\sqrt{3} q_{2,1} & -\sqrt{3} q_{2,-1} & -q_{2,1} & -\sqrt{3}
				q_{2,-2} & -q_{2,0} - \sqrt{3} q_{2,2}
	\end{array}\right)	\ ,
\label{B2}
\end{equation}
and
\begin{eqnarray}
\label{B4}
\tilde B_4 &=& \frac{1}{\sqrt{56}} \times \\
&&\left( \begin{array}{ccccc}
  \sqrt{\frac{2}{5}} q_{4,0} + \sqrt{14} q_{4,4} & \sqrt{14} q_{4,-4} &
\sqrt{6} q_{4,2} &
    q_{4,-1} + \sqrt{7} q_{4,-3} & - q_{4,-1} + \sqrt{7} q_{4,3} \\
  \sqrt{14} q_{4,-4} & \sqrt{\frac{2}{5}} q_{4,0} - \sqrt{14} q_{4,4} &\sqrt{6}
q_{4,-2} &
    - q_{4,-1} - \sqrt{7} q_{4,3} & -q_{4,-1} +  \sqrt{7} q_{4,-3} \\
  \sqrt{6} q_{4,2} & \sqrt{6} q_{4,-2} & \sqrt{\frac{72}{5}} q_{4,0} &
    \sqrt{12} q_{4,-1} & \sqrt{12} q_{4,1} \\
  q_{4,-1} + \sqrt{7} q_{4,-3} & -q_{4,1} - \sqrt{7} q_{4,3} &
    \sqrt{12} q_{4,-1} & -\sqrt{\frac{32}{5}} q_{4,0} - \sqrt{8} q_{4,2} &
    \sqrt{8} q_{4,-2} \\
  - q_{4,1} + \sqrt{7} q_{4,3} & - q_{4,-1} + \sqrt{7} q_{4,-3} &
    \sqrt{12} q_{4,1} & \sqrt{8} q_{4,-2} &
    -\sqrt{\frac{32}{5}} q_{4,0} + \sqrt{8} q_{4,2}
	\end{array}\right)\nonumber
\end{eqnarray}

\subsection{The Born-Oppenheimer energy minima}
\label{BOminima:sect}

The classical (purely static) solution of the JT problem consists in
finding the minimum of the Born-Oppenheimer (BO) potential energy defined
as the sum of the potential term in (\ref{H0}) plus the lowest eigenvalue
of the hole-vibron interaction matrix
\begin{equation}
{\cal H}
\left({\vec q}_2,{\vec q}_4\right)=
g_2\frac{\hbar {\omega}_2}{2}\tilde B_2 +
g_4\frac{\hbar {\omega}_4}{2}\tilde B_4
\label{int_matrix}
\end{equation}
in the (5+9)-dimensional space of the classical $q_{L,m}$ coordinates. This
classical minimum is generally not unique. More typically, there will be
continuous sets of minima, forming well defined manifolds embedded in the
multi-dimensional coordinates space\cite{Ceulemans}. We call one such set
of minima of the adiabatic potential Jahn-Teller Manifold (JTM). The
topology of this manifold crucially affects the possible presence of a
Berry phase. We study the JTM following the approach of \"Opik and
Pryce\cite{OP}, more recently reformulated by Ceulemans in a clear and
useful form\cite{Ceulemans}.

Consider a generic one-fermion state
\begin{equation}
\mid \eta \rangle = \sum_{j=1}^5 \eta_{j} \tilde c_{j,\uparrow}^\dagger
			| 0 \rangle ~ ,
\label{holeket}
\end{equation}
where without loss of generality, the $\eta_j$ can be taken real, with the
normalization constraint
\begin{equation}
\sum_{j=1}^5 {\eta}_j^2 = 1 ~ .
\label{normalization}
\end{equation}
The quantum amplitudes ${\vec \eta}$ live therefore on a $S_4$ sphere in
5-dimensions, which we indicate as the electronic sphere (ES).

Within this notation, it is convenient to rewrite the classical
potential energy as
\begin{equation}
V({\vec q_2},{\vec q_4},{\vec \eta}) = \sum_{L=2,4} \frac{\hbar \omega_L}{2}
	\sum_{l=-L}^{L} q_{L,l}^2 +
	{\vec \eta}^T \cdot {\cal H}({\vec q_2},{\vec q_4})\cdot{\vec \eta}
\label{Veta}
\end{equation}
at fixed distortion ${\vec q_2},{\vec q_4}$ {\em and} electron state ${\vec
\eta}$. Minimization of $V$ with respect to ${\vec \eta}$ yields, by
definition, the (adiabatic) BO potential.  If we keep ${\vec \eta}$ fixed
instead, and minimize (\ref{Veta}) with respect to $\vec q = (\vec q_2
,\vec q_4)$, then we get the extremal points of the BO potential in
configuration space as
\begin{equation}
	\vec{q}=\vec{q}({\vec \eta})
\label{q_of_min}
\end{equation}

Inserting (\ref{q_of_min}) in (\ref{Veta}), we obtain a function $\hat{V}$
of the electronic coordinates alone, hence a function defined on the ES,
not on the space of distortion like the BO potential. We still have to
ensure that the electronic vector is an eigenvector of the interaction
Hamiltonian. Ceulemans\cite{Ceulemans} realized that this condition is
equivalent to a second minimization procedure, on the ES; moreover, he
recognized that the $\hat{V}$ function is isoextremal to the BO surface,
that is, its extremal points on the ES correspond (in a 2:1 way) to
extremal points of the BO surface. Then the JTM has the same symmetry
properties of the set of minima of $\hat{V}$.

We apply this procedure of reverse-order minimization to the present system.
We list here the distortions $q_2$ and $q_4$ which minimize $V$ at fixed
$\vec \eta$ on the ES:

\begin{eqnarray}
q_{2,2} & = & - \frac{2}{\sqrt{7}} g_2 \eta_1 \eta_3 - \frac{1}{2}
\sqrt{\frac{3}{7}}
                g_2 (\eta_4^2 - \eta_5^2) \nonumber \\
q_{2,1} & = &  \sqrt{\frac{3}{7}} g_2 (\eta_2 \eta_4 + \eta_1 \eta_5)
              + \frac{1}{\sqrt{7}} g_2 \eta_3 \eta_5 \nonumber \\
q_{2,0} & = & -\frac{1}{\sqrt{7}} g_2 (\eta_1^2 + \eta_2^2 - \eta_3^2)
              + \frac{1}{2 \sqrt{7}} g_2 (\eta_4^2 + \eta_5^2) \label{q2} \\
q_{2,-1} & = & \sqrt{\frac{3}{7}} g_2 (\eta_2 \eta_5 - \eta_1 \eta_4)
              + \frac{1}{\sqrt{7}} g_2 \eta_3 \eta_4 \nonumber \\
q_{2,-2} & = & - \frac{2}{\sqrt{7}} g_2 \eta_2 \eta_3 + \sqrt{\frac{3}{7}}
                g_2 \eta_4 \eta_5 \nonumber
\end{eqnarray}
\begin{eqnarray}
q_{4,4} & = & - \frac{1}{2} g_4 (\eta_1^2 - \eta_2^2) \nonumber \\
q_{4,3} & = & \frac{1}{\sqrt{2}} g_4 (\eta_2 \eta_4 - \eta_1 \eta_5) \nonumber
\\
q_{4,2} & = & - \sqrt{\frac{3}{7}} g_4 \eta_1 \eta_3 +
              \frac{1}{\sqrt{7}} g_4 (\eta_4^2 - \eta_5^2) \nonumber \\
q_{4,1} & = & \sqrt{\frac{1}{14}} g_4 (\eta_2 \eta_4 + \eta_1 \eta_5)
              - \sqrt{\frac{6}{7}} \eta_3 \eta_5 \nonumber \\
q_{4,0} & = & -  \frac{1}{2 \sqrt{35}} g_4 (\eta_1^2 - \eta_2^2 - 6 \eta_3^2
              +4 \eta_4^2 + 4 \eta_5^2) \label{q4} \\
q_{4,-1} & = & -\sqrt{\frac{1}{14}} g_4 (\eta_1 \eta_4 + \eta_2 \eta_5)
              -  \sqrt{\frac{6}{7}} \eta_3 \eta_4 \nonumber \\
q_{4,-2} & = & - \sqrt{\frac{3}{7}} g_4 \eta_2 \eta_3 +
              \frac{2}{\sqrt{7}} g_4 \eta_4 \eta_5 \nonumber \\
q_{4,-3} & = & -\frac{1}{\sqrt{2}} g_4 (\eta_1 \eta_4 + \eta_2 \eta_5)
\nonumber \\
q_{4,-4} & = & - g_4 \eta_1 \eta_2 \nonumber
\end{eqnarray}
We can choose a (hyper-polar) representation for the electronic coordinates
such that the unitarity condition (\ref{normalization}) is automatically
satisfied
\begin{eqnarray}
\eta_1 & = & \sin \theta_1 \sin \theta_2 \sin \theta_3 \cos \theta_4,
\nonumber\\
\eta_2 & = & \sin \theta_1 \sin \theta_2 \sin \theta_3 \sin \theta_4,
\nonumber\\
\eta_3 & = & \cos \theta_1, \label{etas} \\
\eta_4 & = & \sin \theta_1 \cos \theta_2, \nonumber \\
\eta_5 & = & \sin \theta_1 \sin \theta_2 \cos \theta_3. \nonumber
\end{eqnarray}
Here, $\theta_i$ i=1,2,3 are the azimuthal angles and $\theta_4$ the anomaly.
Substituting (\ref{etas}) in (\ref{q2}) and (\ref{q4}), and then back into
(\ref{Veta}) we obtain an expression for the minimum $E_{JT}$ of
$\hat{V}$:
\begin{equation}
E_{JT} = - \frac{1}{70} (5 g^2_2 \omega_2 + 9 g_4^2 \omega_4).
\label{EJT}
\end{equation}
and for the JTM coordinates
\begin{equation}
|\vec q_2|  = \frac{1}{\sqrt{7}} g_2 \ , \
|\vec q_4| = \frac{3}{\sqrt{35}} g_4 \ .
\label{JTM:eqn}
\end{equation}

\subsection{Is there a Berry phase?}

As Eq. (\ref{EJT}) shows, $E_{JT}$ is independent of the $\eta$ parameters,
and is therefore a {\em constant on the whole} ES. This is a nontrivial
consequence of the large amount of geometrical freedom which is allowed to
the molecule by the large degeneracy of the vibron modes: on the contrary,
when the dimension of $\vec q$-space is less than that of the ES (like for
example in the $h\otimes G$ JT system -- relevant for the coupling of $h_u$
with the $G_g(3)$ mode, which has no $H_g$ partner), $E_{JT}$ is not flat,
but is characterized by some amount of corrugation.  The isoextremality
property of the ${\cal D}^{(2)} \otimes ({\cal D}^{(2)} \oplus {\cal
D}^{(4)})$ implies that the JTM, must have locally the same structure as
the ES, here $S^4$, a sphere in five dimensions.

However, upon closer scrutiny, we note a topological difference between the
ES and the JTM, due to the 2:1 correspondence mentioned above.  In
particular, the ES is a projective space, as opposite points on the sphere
correspond to the same point in the $q$-coordinates space (they belong to
the same ``ray''). This is equivalent to the fact that normalized
eigenvectors of a symmetric matrix are defined up to a sign. The JTM does
not share this property, since there are no pairs of points on the JTM that
generate the same $\vec\eta$. Here is the source of the Berry phase: to
each path in the JTM there corresponds a path on the ES, but the two are
topologically different. In particular, a loop in the JTM may correspond
either to a closed path, or to a path from a point to its opposite
(antipode) on the ES. In general, the latter path cannot be reduced to a point,
and there will be a topological effect (Berry phase).  This fact is
illustrated in Fig.\  \ref{ES:fig} on an ordinary $S_2$ sphere, in
3-dimensional space, representing the ES.
The line $\Gamma$ corresponds to a closed loop on the JTM: there is no way
to make point B coincide with point A in such a way that then we can just
shrink the path and have it coincide to a trivial one such as $\Gamma_1$.
However, we are going to encounter situations where apparently nontrivial
loops become shrinkable to a point {\em via} singularities in the mapping
between the ES and the JTM.

Of the most general set of coupling parameters, we consider three special
cases: (a) ${\cal D}^{(2)} \otimes {\cal D}^{(2)}$, i.e.  $g_2\neq0$, $g_4=
0$; (b) ${\cal D}^{(2)} \otimes {\cal D}^{(4)}$, i.e. $g_2=0$, $g_4\neq0$;
and finally (c) ${\cal D}^{(2)} \otimes ({\cal D}^{(2)} \oplus {\cal
D}^{(4)})$, i.e. $\omega_2=\omega_4$, $g_2=g_4\neq0$.

(a) ${\cal D}^{(2)} \otimes {\cal D}^{(2)}$. Here, the $\vec q- \vec\eta$
mapping is given by Eq.\ (\ref{q2}) and (\ref{etas}). A class of closed
paths on $q$-manifold does, as always, transport $\vec\eta$ into its
antipode $-\vec\eta$. However, we find in this case that a whole
(``equatorial'') line joining two antipodes on the ES, defined by $\theta_1
= \theta_2 =\theta_3 =\pi /2$ (i.e. $\eta_1^2 + \eta_2^2 = 1 ,
\eta_3=\eta_4=\eta_5 = 0$) and parameterized by $\theta_4=0\to\pi$, in fact
corresponds to a {\em single} point ${\vec
q}_2=(0,0,-\frac{g_2}{\sqrt{7}},0,0)$ in $q$-space. The antipodes are
therefore topologically joined, and there are no loops on the JTM that
cannot be deformed to a point. Hence, there is no Berry phase in this case.

(b) ${\cal D}^{(2)} \otimes {\cal D}^{(4)}$. The $\vec q- \vec\eta$ mapping
is now through Eq.\ (\ref{q4}) and (\ref{etas}). A closed loop on the JTM
transports $\vec\eta$ into $-\vec\eta$ if it encircles a conical
intersection in configuration space. This is clearly the case of the line
described in (a), here corresponding to
\begin{eqnarray}
q_{4,4} & = & - \frac{1}{2} g_4 \cos (2 \theta_4) \nonumber \\
q_{4,0} & = & - \frac{1}{2 \sqrt{35}} g_4 \cos (2 \theta_4) \\
q_{4,-4} & = & - \frac{1}{2} g_4 \sin (2 \theta_4) \nonumber
\end{eqnarray}
Now only the $\vec q$ antipodes are topologically equivalent, while the
loops that join them are nontrivial: there is a Berry phase of $\pi$ in the
electronic transport in this case.

(c) ${\cal D}^{(2)} \otimes ({\cal D}^{(2)} \oplus {\cal D}^{(4)})$. In
this case the $\vec q- \vec\eta$ mapping passes through both Eq.\
(\ref{q2}) and Eq.\ (\ref{q4}). The Berry phase of case (b) is therefore
conserved upon addition of the ${\cal D}^{(2)}$ mode. This case has an
instructive analogy to simpler JT-coupled systems.

\subsection{``Propagation'' of Berry phases}

In case (c) above, we add to a Berry-phase entangled DJT system an
additional $L=2$ vibron mode, also coupled to the same electronic state.
It is interesting to consider what happens to the Berry phase when the
added mode, on its own, carries no Berry phase. The enlargement of the
${\vec q}$-space introduces a large freedom for creating new loops in
${\vec q}$. One might in principle think that some of these loops could
manage to avoid the conical intersections of BO levels, thus healing the
connectivity properties of the JTM, and short-circuiting out the Berry
phase. We found instead that this is not the case, and ${\cal D}^{(2)}
\otimes ({\cal D}^{(2)} \oplus {\cal D}^{(4)})$ retains the Berry phase of
${\cal D}^{(2)} \otimes {\cal D}^{(4)}$.

An example of analogous situation is the $t \otimes (T\oplus E)$ problem,
extensively studied in the literature\cite{ob69,ob71}: the $t \otimes T$
problem is Berry-phase--entangled, while the $t\otimes E$ is not. It is
well known that $t \otimes (T\oplus E)$ problem retains the same
topological properties, including the Berry phase of $\pi$, of the $t
\otimes T$ system, the effect of the $E$ mode consisting essentially only
in flattening the JTM, and removing inter-minima barriers.

The reason why new loops seem to be ineffective in destroying the Berry
phase in the total vibron space is that this space is the cartesian product
of the single-mode spaces: the degeneracy points, opening ``holes'' in the
space of the first vibron, propagate (locally, at least) in the global
space along the second-vibron directions. The global conservation of
connectivity in the whole product space is a nontrivial topological
problem, whose general outcome is not known to us.  In both examples above,
we simply note that things go as if the Berry phase carried by the first
mode is strictly conserved.

In the present situation, the coupled ${\cal D}^{(2)}$ vibrational mode
does not present any topological effect (a), whereas the ${\cal D}^{(4)}$,
as seen in case (b), carries a Berry phase. Thus the overall ${\cal
D}^{(2)}\otimes \left({\cal D}^{(2)}\oplus {\cal D}^{(4)}\right)$ system is
affected by topological effects, and this remains true even for different
couplings $g_2\neq g_4\neq0$.

The fullerene cations provide a third example of this behavior. The ${\cal
D}^{(2)}\otimes {\cal D}^{(4)}$ is naturally resolved in its icosahedral
composition $h\otimes G$ plus $h\otimes H$.  In the light of the above
considerations, we expect that in this system the Berry phase should be
carried by the $G$ mode\cite{ob95}.  That this is indeed the case, can thus
be seen as a natural consequence of the absence of Berry phases in
$h\otimes H$ (i.e.\ ${\cal D}^{(2)}\otimes {\cal D}^{(2)}$), and its
presence in ${\cal D}^{(2)}\otimes {\cal D}^{(4)}$.

\section{Semiclassical quantization and the effects of the Berry phase}
\label{semiclass:sect}

So far we have treated the ions as classical. We study here the consequences
of the discussed symmetries of the JTM and of the presence/absence of the
Berry phase on the quantum spectrum of the coupled system, by quantizing
the semiclassical ionic motion. The general form of the
ion kinetic energy is
\begin{equation}
H_0=-\sum_{L=2,4} \frac{\hbar \omega_L}{2} \sum_{l=-L}^{L}
			\frac{\partial^2}{\partial q_{L,l}^2} \ .
\label{KinTot}
\end{equation}
It is convenient to use the classical language for the derivation of the
kinetic operator restricted to the semiclassical region:
\begin{equation}
K   = \frac{1}{2} \sum_{L=2,4} \frac{\hbar} {\omega_L}
		\sum_{l=-L}^{L} \dot{q}_{L,l}^2 \ .
\end{equation}
In the semiclassical limit, valid when $g_L$ are very large, the ionic
motion is confined to the JTM, itself given by Eq. (\ref{JTM:eqn}). Hence,
only the tangential part of the kinetic energy (\ref{KinTot}) is relevant,
which, upon introduction of $\zeta=\frac{\omega_2}{\omega_4}-1$ can be
written
\begin{equation}
	K= \frac{\hbar} {2\omega_2} \sum_{ij}
			\dot{\theta_i} G^{ij} \dot{\theta_j} \ ,
\end{equation}
where $G^{ij}$ is the metric induced on the JTM by the electronic
parametrization:
\begin{equation}
G^{ij} = \sum_{L=2,4} (1+\delta_{L,4}\zeta) \sum_{l=-L}^{L}
{\partial}_{\theta_i} q_{L,l}  {\partial}_{\theta_j} q_{L,l} \ .
\end{equation}
Not much progress is possible in this form, due to the general low-symmetry of
the
kinetic term. However, in the special case of equal frequencies
($\omega_2=\omega_4=\omega$: $\zeta=0$) and equal couplings, where the true
symmetry of the problem is SO(5)\cite{Pooler}, the metric is rewritten as:
\begin{equation}
G^{ij} = {\partial}_{\theta_i} \vec{q} {\partial}_{\theta_j} \vec{q}
\end{equation}
Thus, the kinetic energy corresponds to that of a free particle on a sphere
in 5 dimensions:
\begin{equation}
K = I \frac{\hbar} {\omega_2}
	( {\dot{\theta_1}}^2 + {\sin \theta_1}^2 {\dot{\theta_2}}^2 +
	{\sin \theta_1}^2 {\sin \theta_2}^2 {\dot{\theta_3}}^2 +
	{\sin \theta_1}^2 {\sin \theta_2}^2
	{\sin \theta_3}^2 {\dot{\theta_4}}^2).
\label{ecinetica}
\end{equation}
$I$ is the 'inertial momentum' of the system
\begin{equation}
I = |\vec q|^2 =\frac{2}{5}g^2
\end{equation}
The corresponding Laplace-Beltrami operator (describing the free motion of
a quantum particle on the manifold parametrized by the $\theta_i$ variables)
is given by the formula\cite{Barut}
\begin{equation}
\Delta = \frac{1}{\sqrt{G}} \partial_{\theta_i} G^{ij} \sqrt{G}
	\partial_{\theta_j} ~ ,
\label{laplace}
\end{equation}
where $G$ is the absolute value of the determinant of the metric tensor.
This operator is the usual representation of the quadratic Casimir
operator of SO(5) on the coset manifold SO(5)/SO(4), which is actually the
sphere $S^4$: this is a consequence of the exact SO(5) symmetry of the
equal-coupling case.

The particle moves freely on the JTM, with harmonic oscillations orthogonal
to it. The corresponding energy levels are therefore
\begin{equation}
E = E_{JT} + \frac{{\hbar}^2}{2 I} l(l+3) +
  \hbar {\omega} \sum_{\gamma = 1}^{10} (n_{\gamma} + \frac{1}{2})
\label{levels24}
\end{equation}
$E_{JT}$ is the static JT energy (\ref{EJT}). The last term is due
to the vibrations orthogonal to the JTM. The $l(l+3)$ term corresponds to
the eigenvalues of the Laplace-Beltrami operator on the sphere; $l$ is the
label of the angular hyperharmonics on $S^4$, that realize the
$[l,0]$ representations of SO(5), whose degeneracy is given by
\begin{equation}
N_l = \frac {(2l+3) (l+2) (l+1)}{6}.
\end{equation}
The Berry phase imposes precise selection rules on these levels: since the
one-hole state changes sign after a closed loop around the JTM, and the
overall wave function needs to be single-valued, also the vibrational wave
function has to change sign. Odd-$l$ states have this properties, and these
are the allowed states that survive the Berry phase selection rule. This
implies that the ground state is $l=1$ (fivefold degenerate) and the first
excited state is $l=3$ (30-fold degenerate).  Results of exact
diagonalization on a truncated basis\cite{AMT,Gunnarsson,Reno}
\begin{equation}
\label{psibasis:eqn}
\Psi =\sum
	a_{0\alpha}c_\alpha^\dagger\left|0\right>+
	a_{i\alpha}b_i^\dagger c_\alpha^\dagger\left|0\right>+
	a_{ij\alpha}b_i^\dagger b_j^\dagger c_\alpha^\dagger\left|0\right>+
	...
\end{equation}
shown in Fig.\ \ref{2cross2+4:fig},
confirm the spectral structure described here for the semiclassic limit. At
all values of the coupling $g$, as expected, the degeneracies are those
characteristic of SO(5). We have done also numerical diagonalizations in
the more general case of different coupling and/or frequencies: the
spectra, as expected, display lower -- SO(3) -- symmetry, but a qualitatively
similar behavior.

\subsection{Two holes in spin-singlet state}

In the spin-singlet $n$=2 holes case, $^1$C$_{60}^{2+}$ we find two main
differences with the $n$=1 $^2$C$_{60}^{+}$ case. First, the average
distortion $|\vec q| =2\sqrt{\frac25}g$ is twice as large as in the $n$=1
case, $I$ and the $E_{JT}$ energy being 4 times as large. Secondly, and
most importantly, there is a change in the selection rules. The two holes
live in the same single-hole eigenvector of ${\cal H}({\vec q_2},{\vec
q_4})$, the electronic wave function being the product of two equivalent
eigenfunctions for the space coordinates and an antisymmetric part for the
spin degrees of freedom. Since each hole state changes sign after a closed
loop in the JTM, the overall hole state does not. Thus the vibrational wave
function must also be single-valued in order to have an overall
single-valued wave function for the system. As a consequence, only even-$l$
states survive: in particular, the ground state is $l=0$ (non-degenerate)
and the first excited state is $l=2$ (14-fold degenerate). Exact
diagonalization results, shown in Fig.\ \ref{2cross2+4:fig}(b), confirm that
the symmetry and, to a certain degree the spectral structure described here
for the semiclassic limit are retained in the weak-coupling limit.

\subsection{The 2 $\otimes$ (2 $\oplus$ 4) unequal couplings case}
\label{differc}

In the more general non-equal coupling case $g_2\neq g_4$, including also
the $g_2=0$ case, we could try to use the metric as for the equal coupling
in order to obtain an analogous expression for the kinetic
energy. Unfortunately, when the couplings are different, we do not get a
treatable expression like (\ref{ecinetica}), and we cannot follow the same
route for quantizing the system.

We recall instead that the structure of the JTM is the same as that of ES,
and in particular that SO(5) symmetry is retained. The expression of the
minimum potential energy (\ref{EJT}) suggests a flat JT trough, for any
values of the frequencies and couplings. Therefore, it is likely that the
free motion of the particle on the JTM is described again by the quadratic
Casimir operator of SO(5), retaining for the low-energy excitations a
spectral structure similar to Eq. (\ref{levels24}).

However, although the full SO(5) symmetry is retained in the $g\to\infty$
semiclassical limit, for any non-infinite value of the couplings, the SO(5)
representations will be split into representations of the SO(3) group
\cite{Iachello}. From this point of view we can see the differences between
the frequencies and between the coupling strengths as symmetry-reducing
perturbations, that become effective when moving away from the
strong-coupling limit.

\subsection{The 2 $\otimes$ 2 case}
\label{2cross2}

This case, corresponding to $h_u\otimes H_g$ in a C$_{60}^+$ cation, is
very interesting because of the peculiar features described above, i.e.
absence of topological singularities and, in particular, absence of a Berry
phase.  Furthermore, this case is worth additional analysis, since it
provides an alternate limit of our approach for the different coupling case
(of which this is an extreme case).

We treat again the system as a quantization problem on a five-dimensional
sphere. In this case, it is possible to find the extremal eigenvalues
of the interaction matrix ${\cal H}({\vec q_2})=\frac{1}{2}
g_2\hbar\omega_2 \tilde B_2$, as solutions of the characteristic
polynomial equation for $\tilde B_2$:
\begin{equation}
\lambda^5 - |\vec q_2 |^2 \lambda^3 + 7 f_0(\vec q_2) \lambda^2
+ \frac{12}{49} |\vec q_2 |^4 \lambda - 4 f_0(\vec q_2) |\vec q_2 |^2 = 0
\label{charpol}
\end{equation}
where
\begin{eqnarray}
f_0 (\vec q_2) &=& -\frac{1}{49 \sqrt{7}}
	[3 q_{2,0} (q_{2,1}^2 + q_{2,-1}^2)
	- 6 q_{2,0} (q_{2,2}^2 + q_{2,-2}^2) \nonumber\\
	& &+ 2 q_{2,0}^3 + 3 \sqrt{3} q_{2,2} (q_{2,1}^2 -
	q_{2,-1}^2) - 6 \sqrt{3}
	q_{2,1} q_{2,-1} q_{2,-2} ] \ .
\label{f0}
\end{eqnarray}
This polynomial can be decomposed as
\begin{equation}
(\lambda^2 - \frac{4}{7} |\vec q_2 |^2)
\left[\lambda^3 -\frac{3}{7} |\vec q_2 |^2 \lambda
 + 7 f_0(\vec q_2)\right] = 0.
\end{equation}
The extremal eigenvalues  of $\tilde B_2$ are
\begin{equation}
\lambda_{1/5} =
	\mp \frac{2}{\sqrt{7}} |\vec q_2 | \ ,
\end{equation}
in accord with the SO(5) symmetry of the JTM in the 5-dimensional ${\cal
D}^{(2)}$ space. The three remaining intermediate eigenvalues can be
expressed as
\begin{eqnarray}
\lambda_2&=& |\vec q_2 |
	\frac{2}{\sqrt{7}} \cos\left(\frac{\Psi}{3}-\frac{2\pi}{3}\right)
			\nonumber\\
\lambda_3&=& |\vec q_2 |
	\frac{2}{\sqrt{7}} \cos\left(\frac{\Psi}{3}\right)
			\nonumber\\
\lambda_4&=& |\vec q_2 |
	\frac{2}{\sqrt{7}} \cos\left(\frac{\Psi}{3}+\frac{2\pi}{3}\right)
\end{eqnarray}
where
\begin{equation}
\cos \Psi = -49 \sqrt{7}
        \frac{f_0(\vec q_2)}{2 |\vec q_2 |^3} \ .
\end{equation}
In order to obtain the eigenenergies of ${\cal H}{(\vec q_2)}$ one has
to multiply these eigenvalues by a factor $\frac{1}{2} g_2 \hbar \omega_2$.
By inspection, $\lambda_{2,3,4}$ are bound in the $[\lambda_1,\lambda_5]$
interval. However, it is possible to find many points on the JTM such that
$\lambda_1=\lambda_2$, for example the ``south pole'' $q_0/|\vec q|=-1$ and
a one-dimensional manifold on the ``equator'' $q_0 = 0$. Since
$\lambda_1$ is constant and $\lambda_2$ is analytical in ${\vec q}_2$, in
particular around the tangency points where the two eigenvalues get
degenerate, these degeneracies are not conical intersections of the two
adiabatic surfaces. The two surfaces have therefore just a contact of second
order at these points. The absence of conical intersections implies the absence
of any topological effect\cite{lh}.

Because of the absence of a Berry phase, already discussed in Sect.\
\ref{topology:sect}, the strong coupling behavior of the system, described
in terms of the SO(5) representations, is such that we expect a [0,0] (non
degenerate) ground state, a [1,0] (fivefold degenerate) first excited
state, a [2,0] (14-fold degenerate) second excited state, and so on.
Moreover, absence of topological effects makes the two-holes case not
essentially different from the one-hole problem in the strong coupling
limit. Of course, the static JT energy and inertial momentum is just four
times larger than in the one hole case, but the level sequence in the
strong-coupling limit is the same.

Coming back to the single-hole case, let us now move away from the
strong-coupling limit, to discuss the more general case of an arbitrary
coupling $g_2$. We might na\"ively expect, at first, the ground state
symmetry to be preserved for arbitrary $g_2$, as was the case in other DJT
problems examined previously\cite{AMT}. However, a closer inspection
reveals that this is not true, and rather that there must be a level
crossing as a function of $g_2$. A first indication comes from the fact
that the ground state must in fact be $L=2$, i.e. fivefold degenerate, as
$g_2\to 0$, reflecting just the uncoupled hole ($h_u$) degeneracy. A second
hint comes from the tangency of the lowest with the second BO surface,
noted above. Could this tangency have a role in the expected level crossing
from $L=2$ at $g\to 0$ to $L=0$ at $g\to\infty$?  To study this
possibility, we resort to numerical diagonalization of Hamiltonian
(\ref{Hschematica}) on a truncated basis (\ref{psibasis:eqn}), for finite
values of $g_2$.  As pointed out above, for finite coupling the symmetry of
the system is strictly SO(3). Therefore, we expect the large degeneracies
characteristic of SO(5) symmetry to build up from converging SO(3) states
for large $g_2$. The [0,0] and [1,0] states map directly on the ${\cal
D}^{(0)}$ and ${\cal D}^{(2)}$ SO(3) representations respectively
\cite{Iachello}, while the SO(3) composition of [2,0] is ${\cal
D}^{(2)}\oplus {\cal D}^{(4)}$. Therefore, we seek the convergence of two
such levels in the strong-coupling limit.

The numerical solution of the 1-hole problem (see Fig.\ \ref{2cross2:fig}(a)),
shows indeed the correct order of the levels for large couplings
($g_2\gtrsim 8$), whereas for $g_2\leq~ \sim$8 the [0,0] and [1,0]
levels cross and become inverted.  This level inversion can in fact be
precisely seen as a consequence of the tangency of the two adiabatic
potential surfaces described above. The two BO sheets lie close in a
significant region of configuration space. Treating the upper sheet as a
perturbation for the eigenstates of the lower sheet, the energy of all
these states is pushed up by this correction. The $l=0$ (isotropic) states
feel more efficiently this perturbation than the $l>0$ hyperspherical
harmonics, since they can arrange their nodes so as to avoid the tangency
line. In conclusion, the effect of the tangency is to increase the energy
of the $l=0$ level more than that of $l=1$, so that, for coupling not too
large, $g\gtrsim 8$, the latter is the ground state.

In the 2-holes (spin singlet) case (Fig.\ \ref{2cross2:fig}(b)) this inversion
is not present, the ground state being nondegenerate (${\cal D}^{(0)}$) for
any coupling $g_2$. Again, from a weak-coupling approach, this is a simple
consequence of the order of splitting of the d$^2$-fermion
spin-singlet states upon perturbative JT coupling with a ${\cal D}^{(2)}$
vibration, that is (from lowest to highest) ${\cal D}^{(0)}$, ${\cal
D}^{(4)}$, ${\cal D}^{(2)}$. From the strong-coupling side, the only
observed crossover happens between the ${\cal D}^{(2)}$ and ${\cal
D}^{(4)}$ low excitations above the ground-state. This means that the BO
picture of the two holes in the same single-hole orbital, although
appropriate to the strong-coupling limit, is rather poor for intermediate
JT coupling, when the two fermions manage to arrange in a cooperative state
preserving the global ${\cal D}^{(0)}$ symmetry despite the tangency of the
lowest BO sheets.

\subsection{The  $h_u \otimes G_g$ case}
\label{hcrossG}

Looking at Table \ref{ExperEner:table} we can see that there is a single
$G_g$ left, that does not pair with any $H_g$ mode. We must therefore treat
the coupling of mode $G_g(3)$ according to the true icosahedral
picture. Actually the static $h\otimes G$ JT problem has been treated in
the literature\cite{CeulemansII}. The set of absolute minima of the system
if made up of ten isolated points in the four-dimensional $G_g$
space. Tunnel splitting among these ten BO valleys give rise to vibronic
states of symmetry $A$, $G$, and $H$. In Ref.\ \cite{ob95} the
diagonalization of the corresponding matrix has been performed including
the proper provision for the Berry phase. The result is that the $H$
quintet is the true ground state, followed by the $G$ quartet and the $A$
singlet in the order.  Thus, the Berry phase again acts in such a way to
enforce a fivefold degenerate ground state all the way from weak to strong
coupling, in analogy with the $h \otimes (G\oplus H)$ JT problem, and in
contrast to the $h\otimes H$ case discussed in Sect.\ \ref{2cross2}.  This
confirms that, as anticipated, the Berry phase in the $h \otimes (G\oplus
H)$ JT problem comes from $h\otimes G$ sector.

\section{Discussion}
\label{discussion:sect}

We have studied the symmetry aspects and the possible presence of a Berry
phase in C$_{60}^+$ and C$_{60}^{2+}$ in the linear and spherical
approximation. We find that while $^1$C$_{60}^{2+}$ has no Berry phase,
$^2$C$_{60}^{+}$ does, and is very rich and intriguing.

Clearly, the present analysis is only a first step.  In reality, fullerene
is not a sphere, but a discrete icosahedron, and the ${\cal D}^{(4)}$ modes
are split into $H_g \oplus G_g$ modes. Nonetheless the real shape of the
molecule can be seen as a perturbation on the spherical picture (Tab.\
\ref{ExperEner:table}): the SO(5) and SO(3) levels are split but the
selection rules are still present. The \"Opik-Pryce approach can still be
applied in order to obtain the JTM in icosahedral symmetry\cite{CeulemansII}.
Quite generally the BO potential will warp in such a
way to originate a discrete set of minima, instead of a continuous
4-dimensional manifold. The molecular system will execute tunneling among
these minima. However, as shown by Ham\cite{ham}, the Berry phase imposes
the degeneracy of the ground state (irrespective of the presence and height
of barriers between the BO minima), which still is fivefold degenerate
(${\cal D}^{(2)} \rightarrow H$).  Therefore, we expect the ground state
symmetry to be $^2H_u$ for real C$_{60}^+$, which has a Berry phase, and
$^1A_g$ for singlet C$_{60}^{2+}$, which does not.

A quantitative study of the properties of C$_{60}^+$ is clearly beyond the
scope of the present work. In order to accomplish that, a full set of e-v
couplings $g_{H_g(i)}\ i=1,... 8$ and $g_{G_g(i)}\ i=1,... 6$ will have to
be known (either from calculations, or fit to experiments), and inserted
into a diagonalization procedure similar to that previously carried out in
the case of C$_{60}^{n-}$\cite{AMT,Gunnarsson,Reno}. While that remains to
be done, we can list, for the time being, some tentative physical
implications of our results.

(i) {\em Low-energy electron detachment}. In a way analogous to that proven
for the case of  C$_{60}^{-}$, one can surmize that C$_{60}^+$ cannot have any
$A_g$ low-energy states. It follows that if an electron is detached from
C$_{60}$ into a low-energy state, the s-wave cross-section will be
zero. and there will be  a centrifugal barrier in detachment too, similar
in nature to that pointed out for attachment\cite{attachment}. This
effect could be related to the thermoionic anomalies pointed out by
Yeretzian {\em et al.}\cite{Yeretzian}

(ii) {\em Low-energy tunneling excitation}. In the spherical approximation,
the lowest excitation of C$_{60}$ is a 30-fold degenerate SO(5) ``$l=3$''
tunneling multiplet, as required by the Berry phase. Split, as it will be
by the lower SO(3) symmetry, as well as by the true ${\cal I}_h$ effects,
this multiplicity might still be identifiable spectroscopically.

(iii) {\em C$_{60}$ photoemission spectrum}. In the valence photoemission
spectrum of C$_{60}\to$ C$_{60} + e^-_{\vec k}$ the electron kinetic energy
distribution contains very direct information -- in the so-called final
state approximation -- about vibron and hole-vibron coupling in
C$_{60}^+$. The outgoing electron shakes up multi-vibron excitations of
C$_{60}^+$, with a relative probability which can be calculated in
principle, in a manner similar to that used by Gunnarsson\cite{Gunnarsson}
for C$_{60}^-$. Very recent data\cite{bruhwiler} indicate that these
shake-ups do exist and are in fact stronger for C$_{60}$ than for
C$_{60}^-$.

We hope to return to these topics in subsequent work.

\section*{Acknowledgments}
P. De Los Rios is grateful to S. Panzeri for many useful discussions.  We
acknowledge exchange of information and discussion with P. Br\"uhwiler and
F. Iachello. This work was partly sponsored by NATO, through CRG 920828,
and by EEC through ERBCHRXCT 940438.


\begin{thebibliography}{10}
\bibitem{SO5:nota}
{ Of course, this very high symmetry breaks down as soon as the ${\cal
  D}^{(2)}$ and ${\cal D}^{(4)}$ modes are no more degenerate and/or equally
  coupled, reducing the symmetry to the expected SO(3)}.
\end{thebibliography}

\begin{figure}[t]
\caption{ A three-dimensional representation of the (truly
five-dimensional) electronic sphere (ES): the path $\Gamma$ connects opposite
points (antipodes), which correspond to the {\em same} point on the
JTM. Thus $\Gamma$ maps, through Eqs.\  (\protect\ref{q2}) and
(\protect\ref{q4}), on a closed loop in the JTM. This class of loops is
usually entangled with a nonzero Berry phase, while trivial loops, such as
$\Gamma_1$ can be smoothly deformed to one point, with associated zero
Berry phase. However, in the case $g_4=0$, due to the singularity of
mapping (\protect\ref{q2}), a whole equatorial line $\Lambda$ maps to a
single ${\vec q}$-point. In this special case, point B identifies with
point A on the JTM, thus $\Gamma$ becomes equivalent to $\Gamma_1$, and no
Berry phase is possible.
\label{ES:fig}}
\end{figure}

\begin{figure}[t]
\caption{The low-lying states of the spectrum of the $L=2\otimes (L=2\oplus
L=4)$ ([1,0]$\otimes$[2,0], in SO(5) notation) JT system, as a function of
the coupling $g=g_2^2=g_4^2$. Symmetry labels are indicated according to
the SO(5) group representations. The truncated basis includes up to 7
vibrons. Panel (a): the levels originated from the 0- and 1-vibron states,
for $n=1$ holes (see also Table \protect\ref{PerturbCoef:table} for the
small-$g$ limit). Panel (b): the three levels originating from the
0-vibrons multiplet for $n=2$ holes. The tenfold-degenerate [0,2] orbital
state is in a triplet spin configuration.
\label{2cross2+4:fig}}
\end{figure}

\begin{figure}[t]
\caption{The low-lying states of the spectrum of the $L=2\otimes L=2$
coupled JT system as a function of the coupling $g_2^2$. We include in the
diagonalization states with up to 35 vibrons. For $n=1$ fermions included
(panel a), the reference level is taken as the lowest $L=2$ state, that is
also the ground state for $g_2<\sim8$.  For $n=2$ fermions included
(panel b) the energies are excitations above the $L=0$ ground state. Only
spin-singlet states are plotted.
\label{2cross2:fig}}
\end{figure}

\begin{table}
\begin{center}
$
\begin{array}{cccc}
\hline
\hline
\begin{array}{c}
\text{SO(3)} \\
L
\end{array}
& {\cal I}_h &
\begin{array}{c}
\text{Calculated} \cite{CeulemansIII} \\
\text{cm}^{-1}
\end{array} &
\begin{array}{c}
\text{Calculated} \cite{koh} \\
\text{cm}^{-1}
\end{array}\\
\hline
\hline
2 &
\begin{array}{c}
H_g (1)\\
H_g (4)
\end{array}
&
\begin{array}{c}
214 \\
727
\end{array}
&
\begin{array}{c}
261 \\
775
\end{array}
\\
\hline
3 & G_g (3) & 557 & 594 \\
\hline
4 &
\begin{array}{c}
\begin{array}{c}
H_g (2)\\
G_g (1)
\end{array}
\\
\hline
\begin{array}{c}
H_g (6)\\
G_g (4)
\end{array}
\end{array}
&
\begin{array}{c}
\begin{array}{c}
387 \\
374
\end{array}
\\
\hline
\begin{array}{c}
1180 \\
774
\end{array}
\end{array}
&
\begin{array}{c}
\begin{array}{c}
435 \\
482
\end{array}
\\
\hline
\begin{array}{c}
1208 \\
1047
\end{array}
\end{array}
\\
\hline
5 & H_g (5)& 1091 & 1098 \\
\hline
6 &
\begin{array}{c}
\begin{array}{c}
H_g (3)\\
G_g (3)
\end{array}
\\
\hline
\begin{array}{c}
H_g (7)\\
G_g (5)
\end{array}
\end{array}
&
\begin{array}{c}
\begin{array}{c}
516 \\
623
\end{array}
\\
\hline
\begin{array}{c}
1477 \\
1412
\end{array}
\end{array}
&
\begin{array}{c}
\begin{array}{c}
730 \\
781
\end{array}
\\
\hline
\begin{array}{c}
1394 \\
1314
\end{array}
\end{array}
\\
\hline
7 &
\begin{array}{c}
H_g (8)\\
G_g (6)
\end{array}
&
\begin{array}{c}
1617 \\
1694
\end{array}
&
\begin{array}{c}
1573 \\
1479
\end{array}\\
\hline
\end{array}
$
\end{center}
\caption{Experimental energies of the JT-active $G_g$ and $H_g$ modes,
organized according to their spherical
parentage\protect\cite{CeulemansIII}. The numbers in parenthesis in the
second column are the position of the mode in order of increasing energies.
\label{ExperEner:table}}
\end{table}

\begin{table}[t]
\begin{center}
$
\begin{array}{ccccc}
\hline
\hline
\begin{array}{c} \text{SO(3)} \\ L \end{array}
& \frac{\Delta E}{g_2^2} & \frac{\Delta E}{g_4^2}
& \frac{\Delta E}{g^2} &
\begin{array}{c} \text{SO(5)} \\ \lbrack l,0] \end{array}
\\
\hline
\hline
2^{(GS)} & -\frac 14 & -\frac 9{20} & -\frac 7{10} & [1,0] \\
\begin{array}{c} 0 \\ 3^{\dagger } \\ 4^{\ddagger } \\ 6
\end{array} &
\begin{array}{c} -\frac 12 \\ -\frac{11}{28} \\ -\frac 9{28} \\ -\frac 14
\end{array} &
\left.
\begin{array}{c} -\frac 9{20} \\ -\frac{121}{280} \\ -\frac{141}{280} \\
-\frac 7{10} \end{array}
\right\}
& -\frac{19}{20} & [3,0] \\
\begin{array}{c} 1 \\ 2^{*} \\ 3^{\dagger } \\ 4^{\ddagger } \\ 5
\end{array} &
\begin{array}{c} -\frac 18 \\ -\frac 14 \\ -\frac 14 \\ -\frac 14 \\ -\frac 14
\end{array} &
\left.
\begin{array}{c} -\frac 9{20} \\ -\frac 9{20} \\ -\frac 9{20} \\ -\frac 9{20}
\\ -\frac{13}{40} \end{array}
\right\}
& -\frac{23}{40} & [1,2] \\
2^{*} & +\frac 3{56} & -\frac 1{280} & -\frac 3{40} & [1,0]\\
\hline
\end{array}
$
\end{center}
\caption{The coefficients of shifts of the vibronic levels for second-order
perturbation in the couplings $g_2$ and $g_4$, for the 0- and 1-vibron
multiplets of levels, for $n=1$ holes. In the first column we list the
labels of the states according to spherical symmetry; the second and third
column report the results of applying as a perturbation $H_{h-v}^{(2)}$ and
$H_{h-v}^{(4)}$ alone respectively; in the next column the perturbation is
done with equal couplings $g_2=g_4=g$. In this last case the true symmetry
of the system is SO(5), that we evidence with the labels in the last
column.  The states marked with $^\dagger$ and $^\ddagger$, are multiply
present in the 1-vibron multiplet, and therefore admix differently with the
three different perturbations. Results in units of
$\hbar\omega_2=\hbar\omega_4$.
\label{PerturbCoef:table}}
\end{table}

\end{document}